\documentclass[journal]{vgtc}                     

\onlineid{0}

\vgtccategory{Research}

\vgtcpapertype{please specify}

\title{Interactive Counterfactual Exploration of Algorithmic Harms in Recommender Systems}

\author{%
  Yongsu Ahn,
  Quinn K Wolter,
  Jonilyn Dick, 
  Janet Dick, and
  Yu-Ru Lin
}

\authorfooter{
  \item
  	Yongsu Ahn is with the University of Pittsburgh.
  	E-mail: yongsu.ahn@pitt.edu

  \item Quinn K Wolter is with the University of Pittsburgh.
  	E-mail: QuinnKWolter@gmail.com.

    \item
  	Jonilyn Dick is with Quest Diagnostics.
  	E-mail: jonilyndick@gmail.com.

   \item
  	Janet Dick is with Quest Diagnostics.
  	E-mail: janetandick@gmail.com.

   \item
  	Yu-Ru Lin is with the University of Pittsburgh.
  	E-mail: yurulin@pitt.edu.
}

\abstract{%
Recommender systems have become integral to digital experiences, shaping user interactions and preferences across various platforms. Despite their widespread use, these systems often suffer from algorithmic biases that can lead to unfair and unsatisfactory user experiences. This study introduces an interactive tool designed to help users comprehend and explore the impacts of algorithmic harms in recommender systems. By leveraging visualizations, counterfactual explanations, and interactive modules, the tool allows users to investigate how biases such as miscalibration, stereotypes, and filter bubbles affect their recommendations. Informed by in-depth user interviews, this tool benefits both general users and researchers by increasing transparency and offering personalized impact assessments, ultimately fostering a better understanding of algorithmic biases and contributing to more equitable recommendation outcomes. This work provides valuable insights for future research and practical applications in mitigating bias and enhancing fairness in machine learning algorithms.
}
\vspace{-2em}

\teaser{
  \centering
  \includegraphics[width=0.85\linewidth]{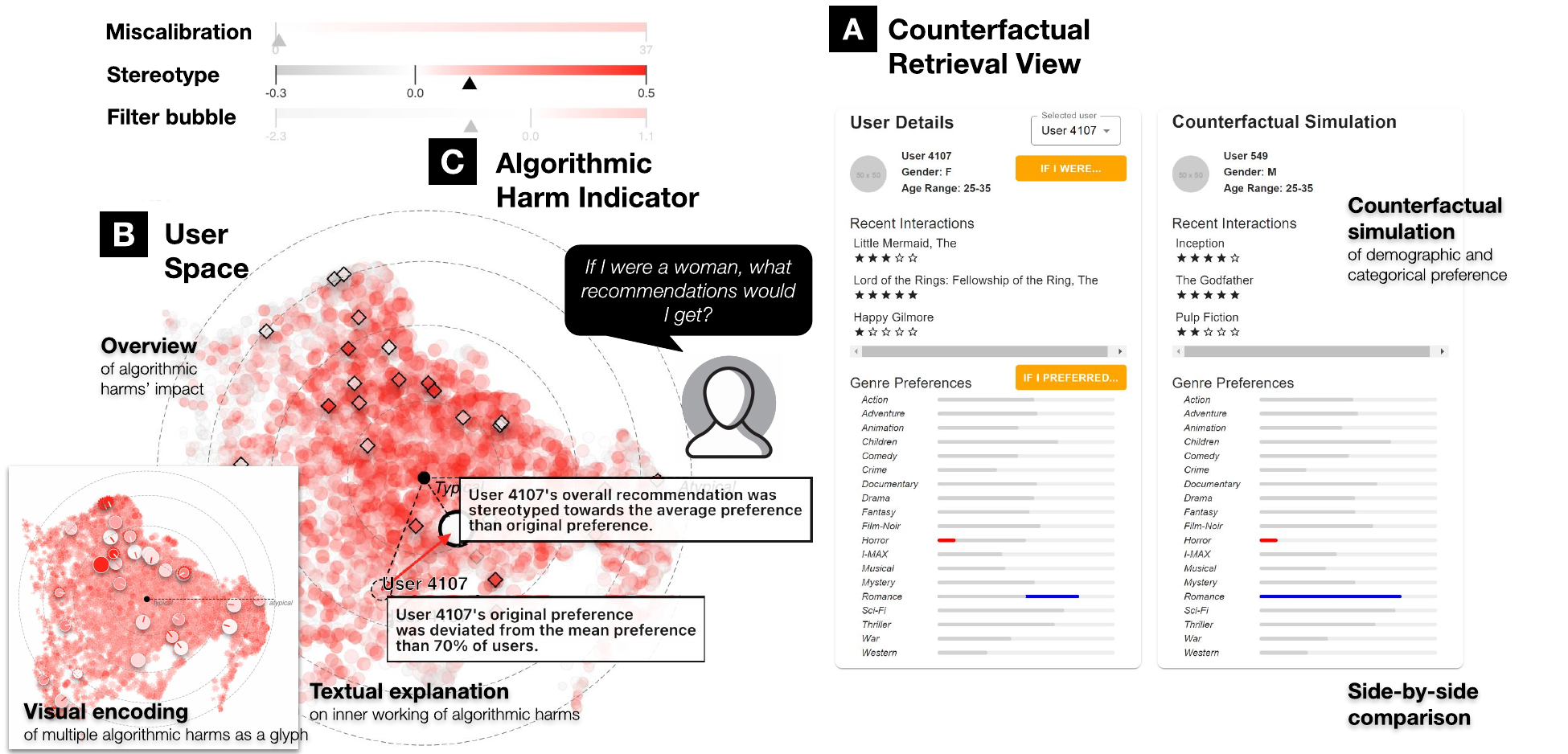}
  \vspace{-0.5em}
  \caption{%
  Interactive dashboard for exploring algorithmic harms in recommender systems. (A) In the Counterfactual Retrieval View, users can explore these "what-if" simulations through altering demographic information or genre preferences. (B) User Space represents users within a high-dimensional preference space, with the selected user and counterfactual user highlighted. The visualization shows how typical or atypical users are grouped and their algorithmic harms represented through glyphs. The system also provide the user with a detailed explanation of the algorithmic harms identified. (C) Users are also provided with visual metrics to show the impact of diverse algorithmic harms on the selected user's recommendations.%
  }
  \vspace{-0.75em}
  \label{fig:teaser}
}

\renewcommand{\manuscriptnotetxt}{}

\graphicspath{{figs/}{figures/}{pictures/}{images/}{./}} 

\usepackage{tabu}                      
\usepackage{booktabs}                  
\usepackage{lipsum}                    
\usepackage{mwe}                       
\usepackage{wrapfig}

\usepackage{mathptmx}                  

\UseRawInputEncoding
\begin{document}



\maketitle
\vspace{-2em}
\section{Introduction}\label{sec:introduction}
The integration of recommender systems into various aspects of digital life has revolutionized user experiences, from personalized content delivery to targeted advertisements. However, these systems, often shrouded in opacity, pose significant challenges related to fairness and bias. As recommender systems increasingly influence users' choices, it becomes crucial to address how algorithmic biases impact decision-making and user satisfaction. 

Despite the algorithmic harms' potential impact towards thousands of users who have different characteristics and perceptions, the solutions are not often human-centered, still largely relying on automatic methods to quantify and mitigate algorithmic harms at one-size-fit-all manner. Our preliminary results indicate that users generally lack the understanding of algorithmic harms, including their mechanisms and the extent to which they impacts recommendations. In addition, perceived impacts vary significantly, highlighting the needs for a nuanced solution to effectively mitigate the algorithmic harms.

Based on the findings revealed from pilot study, we propose an interactive dashboard as a human-centered solution to support users in explore algorithmic harms in recommender systems. By introducing several modules and visualization, we pursue a user-centric exploration of three algorithmic harms, including miscalibration, stereotype, and filter bubble, perceived as exhibiting high distrust and desire to transparency. The current challenges compiled from pilot study are translated into system requirements, textual and visual explanations of algorithmic harms' inner workings and impact, social contextualization, and counterfactual retrieval.

This study aims to bridge the gap by presenting an interactive tool designed to enhance users' understanding of algorithmic harms in recommender systems. Our research not only contributes to the ongoing discourse on algorithmic fairness but also offers actionable insights into how users can engage with and influence the fairness of recommender systems. Through this approach, we seek to foster a more transparent and equitable digital environment, where users are better informed and more actively involved in addressing the biases that affect them.

The contribution of our study includes:

\begin{itemize}
    \item Our study conveys key findings of user perceptions on algorithmic harms in recommender systems. 
    \vspace{-0.25em}
    \item In light of the challenges, we propose a interactive dashboard as a user-centered solution for exploring algorithmic harms with a suite of explanatory modules.
    \vspace{-0.25em}
    \item We introduce a novel design of glyphs to visually summarizing multiple harms.  
\end{itemize}
\vspace{-0.5em}
\section{Related Work}
\label{sec:related-work}

\subsection{Visual analytics for fairness and algorithmic harms}
\label{sec:related-work-1}

Recent studies in visual analytics have advanced tools and visualizations to enable domain users and practitioners to foster fair data-driven decision-making practices. Compared to data-driven measures and methods, human-centered solutions that incorporate visual representations of biases in classification and ranking results have proven effective in helping users make sense of how biases arise \cite{ahn2020fairsight, wexler2019thewhatiftool, bellamy2018aifairness360, wang2023visualanalysis} or specific types of biases, such as intersectional biases in subgroups \cite{cabrera2019fairvis, liu2023faircompass} and biases in image classification \cite{kwon2022dash}. Recently, such systems were developed to be oriented towards involving end-users’ perceptions of fairness in the loop \cite{nakao2022towardsinvolving}and operationalize the real-world decision-making process in specific teams and organizations with the support of decision diagram \cite{liu2023faircompass}. Other studies focus on specialized analyses like graph mining \cite{rissaki2022biascope, xie2022fairrankvis} and causal analysis, employing causal diagram visualizations to promote fairness \cite{ghai2022dbias, yan2020silva}.

Despite these advancements, algorithmic harms and fairness issues in recommender systems, affecting thousands of users, remain largely unaddressed beyond conceptual frameworks for fair recommender systems, which include explanation, audit, and mitigation engines \cite{giannopoulous2021interactivity}. Our study represents a pioneering effort to inform users about various types of algorithmic harms and empower them to comprehend and simulate the impacts within recommender system usage.

\subsection{Interactive counterfactual fairness}
\label{sec:related-work-2}
Recommender systems are known to perpetuate various types of algorithmic harms and effects. While a number of automatic methods to quantify and mitigate such biases have been proposed, studies \cite{ngo2020exploringmentalmodels,baracskay2022diversity,wang2021bias,wang2021useracceptance} indicate that their impacts can vary significantly among different users. For instance, in career recommendations \cite{wang2021bias}, users tend not to perceive gender stereotypes as harmful as long as recommendations are effective for them. \cite{ngo2020exploringmentalmodels} has shown that users develop their own mental models very differently based on experiences and background, resulting in different perception of algorithmic harms. 

In response to the opaque nature of recommender systems, studies have attempted to design various types of explanations such as why/why-not \cite{wilkinson2021whyorwhynot} or personal explanations \cite{kunkel2019letmeexplain} to provide users with rationales behind recommendations. Counterfactual explanations have recently gained attention for their intuitiveness and low cognitive load to understand its perceived efficacy and mitigate filter bubbles \cite{shang2022whyami, cheng2020dece}.

Our study, informed by in-depth interviews with a diverse group of users, introduces an interactive tool designed to help users make sense of algorithmic harms. We present a suite of visualizations, counterfactual explanations, and interactive modules to allow users to make sense of the impact of algorithmic harms over their use of recommender systems.

\section{From User Perception interviews to System Requirements}
\label{sec:pilot-interview}

\subsection{Pilot Interviews}
Our study was motivated by pilot interviews aimed at understanding how current users in existing recommender platforms perceive different types of algorithmic harms. This preliminary study consisted of 1-hour interview sessions with 8 participants who daily consume recommender platforms and content including video (Youtube), social media (Instagram, Facebook), and online news (Google, Bing, and algorithm-based news platforms such as Ground news). Participants were recruited in a diverse pool of nationalities (US, Portugal, Mexico, South Korea, and Taiwan), gender (Male: 2, Female: 2, LGBTQ+: 3), age (18-24: 2, 25-34: 3, 35-44: 2, 45-54: 1), and education levels based on the highest degree completed (High school: 2, Undergraduate: 2, Master's: 3, Ph.D: 1). Each session began with a semi-structured interview allowing them to share their daily use and unsatisfactory experiences of recommendation platforms, followed by card sorting activities that express the degree to which they perceive different types of algorithmic harms regarding trust (i.e., how much do you think it makes you distrust the system if it exists?) or transparent (i.e., how important is it to know if the system informs you of it?). 
In the activity, participants were instructed to place multiple cards describing well-known algorithmic harms including filter bubble, stereotype, popularity bias, miscalibration, and echo chamber within a two-dimensional space, with each axis representing degrees of transparency and trustworthiness, ranging from low to high.

After the coding analysis of pilot studies, we identified three major challenges in user perception of algorithmic harms in recommender systems:\\

\vspace{-0.5em}
\hangindent=2em \textbf{C1. Lack of understanding in algorithmic harms:} Participants had noticeably little understanding of how algorithmic harms work or potentially worsen their experiences. Especially, all pilot interviewees were initially unaware of stereotyping effects, though they considered these serious once informed. \\\vspace{-5pt}

\hangindent=2em  \textbf{C2. Users' focus on categorical preferences.} Despite recommendation studies emphasizing item-level performance, users' interests lie at topic and category levels. Participants mentioned broad categories like health, music, or politics. Three users were particularly interested in how recommendations are distributed and miscalibrated across their interest categories. \\\vspace{-5pt}

\hangindent=2em \textbf{C3. Different levels of perceived impacts over algorithmic harms.} Card sorting results showed that perceived impacts of algorithmic effects vary among users. While all harms were considered harmful by at least one participant, miscalibration, stereotype, and filter bubble triggered higher distrust and desire for transparency. \\\vspace{-5pt}

\hangindent=2em \textbf{C4. Recommendations as social space and needs of transparency:} Five participants viewed recommendations from a social perspective, expressing interest in investigating their groups (e.g., women, users with similar preferences) or identifying themselves within the broader user context. This stems from the perceived nature of recommender systems as spaces where numerous users interaction and collective preferences highly influence the inner workings of algorithms.

\subsection{System Requirements}
Based on the findings from the pilot interviews, we derived the following system requirements to address the identified challenges:\\

\vspace{-0.5em}
\hangindent=2em \textbf{R1. Transparency of algorithmic harms}: To address C1, the system should provide clear and understandable explanations of different algorithmic harms, such as miscalibration, stereotype, and filter bubble effects. These explanations should be accessible to users with varying levels of technical knowledge.\\\vspace{-5pt}

\hangindent=2em \textbf{R2. Category-Level analysis}: To address C2, the system should allow users to analyze and understand recommendations and algorithmic effects at a categorical level. Users should be able to view and compare recommendation behavior across different demographic categories and genre distributions.\\\vspace{-5pt}

\hangindent=2em \textbf{R3. Evaluation of personalized impact of algorithmic harms}: In light of C3, the system should enable users to assess the perceived impact of algorithmic harms on their personal experience using recommendation algorithms. This feature should support individual-level analysis, allowing users to see how specific harms affect them differently compared to other users.\\\vspace{-5pt}

\hangindent=2em \textbf{R4. Social Contextualization}: To address C4, the system should incorporate features that allow users to understand their recommendations within a broader social context. This includes tools to compare their preferences and algorithmic treatment with those of similar demographic groups or broader user bases.

\section{Interactive dashboard for exploring algorithmic harms in recommender systems.}
\label{sec:system}
\subsection{Method}
\label{sec:method}

\subsubsection{Dataset and Algorithm}
In the study, we use the MovieLens 1M dataset that contains demographic attributes such as gender and age. We binarize 1,000,209 user ratings in the dataset to convert it to implicit feedback by dropping all ratings less than 4. ranging from 1 to 5 over movie items across 18 movie genres. We removed all users with less than 20 interactions to ensure the visualization of each user with a meaningful number of interactions. After the preprocessing steps, the data includes 562,800 interactions from 5,180 users and 3,526 items. The dataset includes 18 movie genres to allow us to engage category-level analysis. 

Similar to previous literature \cite{abdollahpouri2020connection, steck2018calibrated}, the implicit feedback data was split into 80\% and 20\% for training and testing using user-fixed setting where the interactions within each user were divided into training and test set based on chronological order. We trained Bayesian Personalized Ranking (BPR), one of recommendation algorithms widely studied in recommendation studies.

\subsubsection{Measures for algorithmic harms}\label{sec:measures}
In this section, we introduce measures for three algorithmic harms, miscalibration, stereotype, and filter bubble, proposed in existing literature.

\textbf{Miscalibration.} Introduced by \cite{steck2018calibrated}, miscalibration quantifies the discrepancy between a user's actual preference $p(\vec{c}|u)$ and predicted preference $q(\vec{c}|u)$ over item categories $\vec{c}$ (e.g., movie genres) for a user $u$.

The individual-level miscalibration, $MC_u$, for a user $u$, uses Kullback-Lieber divergence $D_{KL}$ to calculate the discrepancy between two probability distributions to measure how $q$ deviates from $p$ \cite{steck2018calibrated}, as follows: 

\vspace{-0.25em}
\begin{equation}
    MC_u(p, q) = D_{KL}(p||q) = \sum_{c}{p(c|u)} \log \frac{p(c|u)}{q(c|u)}
    \label{eq:miscalibration}
\end{equation}
\vspace{-0.25em}

Miscalibration ranges from $[0, \inf)$, with smaller values indicating greater similarity between $q$ and $p$. Zero miscalibration represents perfect calibration. System-level miscalibration is denoted as $MC(P, Q)$, where $P$ and $Q$ are sets of actual and predicted preferences for all users.

\textbf{Stereotype.} Following \cite{ahn2024break}, stereotype quantifies the degree of systematic over generalization of a predicted preference $q$ for a user $u$ based on certain characteristics of individuals, which can be computed as follows:

\vspace{-0.5em}
\begin{equation}
    ST_u(p,q) = D_{JS}(p||\bar{P}) - D_{JS}(q||\bar{Q}).
    \label{eq:indi-st}
\end{equation}

where $D_{JS}(p||\bar{P})=\frac{D_{KL}(p||\bar{P})+D_{KL}(\bar{P}||p)}{2}$ is a symmetric measure of quantifying the distance between two probabilistic distributions.

\textbf{Filter bubble.} The term, ``filter bubble'', as defined by Pariser \cite{rowland2011filter}, describes the systematic reinforcement of narrowing exposure to diverse content. It involves entropy as a measure of content diversity where $Entropy(u) = -\sum_{c \in \vec{c}} p(\vec{c}|u) \log p(\vec{c}|u)$. In this study, we quantify the difference in entropy of category-wise preferences between actual ($p$) and predicted ($q$) preferences:

\begin{equation}
    FilterBubble_u(p, q) = DV_u(q) - DV_u(p)
\end{equation}

where $DV_u = Entropy(u) = -\sum_{c \in \vec{c}} p(\vec{c}|u) \log p(\vec{c}|u)$ measures the diversity of a user $u$. We also introduce the term of ''inflated diversity'' \cite{ahn2024break} to describe a counterpart effect of filter bubble in which a user receives an over-diverse prediction $q$.

\subsubsection{Counterfactual retrieval}\label{sec:counterfactual-retrieval}
Our study leverages counterfactual retrieval to support the simulation of “what if I were another user?”. Unlike counterfactual generation, our module is operationalized to retrieve actual users rather than generate them for users to be more immersive of what-if situation. To facilitate the counterfactual search (e.g., what if I were a man?), we borrow matching techniques, which retrieve counterpart users or groups in observational data (i.e., control group) who have as the most similar conditions and characteristics (e.g., categorical preferences) for all variables but a treatment variable (e.g., gender). Our study supports two types of counterfactual queries:

\begin{itemize}
    \item \textbf{Demographic counterfactual}: e.g., if I were a man, what recommendations would I get?
    \item \textbf{Preference counterfactual}: e.g., if I preferred more Sci-fi genre movies, what items would I get? 
\end{itemize}

\subsection{Interactive System}
\label{sec:visualization}

\subsubsection{Counterfactual Retrieval View}
This component provides an interface for counterfactual user retrieval. Users can select demographic or preference-based counterfactual queries to see how different attributes would change their recommendations. For example, they can explore what recommendations they would receive if they were of a different gender or had different genre preferences. By allowing queries based on both genre preferences and demographics, users can identify effects on both item-level and category-level preferences. This view enhances transparency and offers a personalized impact assessment, helping users understand the algorithmic harms they may be facing.

\subsubsection{User Space}
The User Space visualizes users to help them understand the system as a social space in which algorithmic harms may occur. In this two-dimensional visualization users are represented as a glyph or circle whose high-dimensional representations of their category-wise preferences are projected into 2D space using UMAP \cite{mcinnes2018umap}. 


\begin{wrapfigure}{l}{0.55\columnwidth}
\vspace{-1em}
\includegraphics[width=0.6\columnwidth]{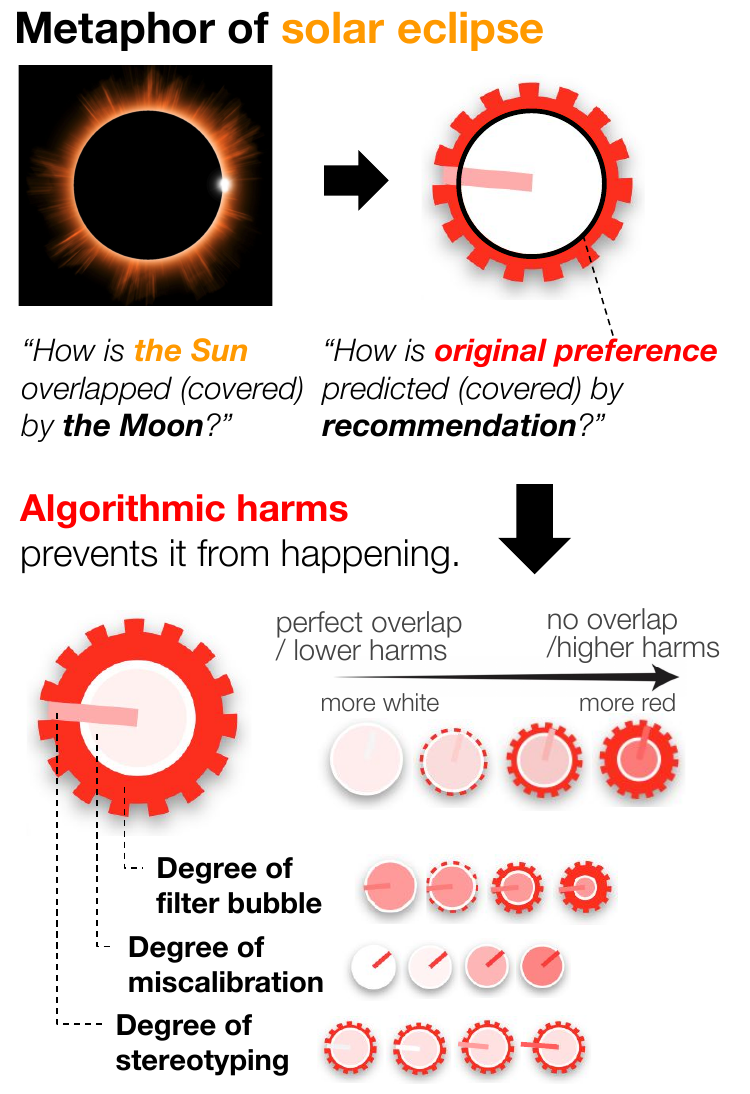}
  \vspace{-1.5em}
  \caption{The details of glyph design for visualizing algorithmic harms.}
  \label{fig:glyph}
  \vspace{-1.2em}
\end{wrapfigure}

\textbf{Glyphs for algorithmic harms.} Given potential impacts of different types of algorithmic harms in recommender systems, we represent users with a glyph to encode multiple harms into a single visual element. For the glyph, we use the metaphor of a solar eclipse: the Sun partially or totally overlapped by the Moon (Fig. \ref{fig:glyph}). Considering a recommender system's objectives, an ideal recommendation algorithm perfectly predicts (i.e., cover) the original preference (i.e., the Sun) with its recommendation (i.e., the Moon), akin to a total solar eclipse. When algorithmic harms occur, they result in imperfect predictions, analogous to a partial or no solar eclipse. To indicate the degree of overall algorithmic harms, the Sun and the Moon are colored red and white, respectively. A greater amount of red in the glyphs indicates a higher degree of algorithmic harm.

A glyph is a composite of visual indicators of three algorithmic harms, translating their mechanisms into intuitive visual metaphors. First, the filter bubble is represented by the thickness of the red outer circle indicating how the original preference (depicted as a Sun-shaped icon, with its size representing the diversity of original preference) is not covered by recommendations (illustrated by an inner circle, with its size indicating the predicted preferences and their diversity). Miscalibration is conveyed through the color of inner circle, indicating the quality of the predicted preference. Lastly, the stereotyping effect is symbolized by a line pointing towards the mean preference. Its red color indicates a higher degree of stereotypes.

\textbf{Prototypical users.} To mitigate cognitive load due to a sheer amount of users presented in the system, we present prototypical users as representatives of primary user characteristics. We use k-medoids clustering \cite{schubert2021fast} to identify actual users as prototypes so that users can be more immersive of ``what-if'' simulation with actual users as counterfactual cases.

\textbf{Visualizing single algorithmic harm or all at once.} While glyphs provide a visual understanding of multiple algorithmic harms at once, users can select one of them listed in the Algorithmic Harm Indicator for a simpler visualization, where users and prototypes are represented with circles and diamond-shaped icons with its color indicating the degree of the selected algorithmic harms.

\subsubsection{Algorithmic Harm Indicator}
The Algorithmic Harm Indicator dashboard provides detailed visualizations and explanations of the three algorithmic harms identified as the largest contributors to mistrust. Users can interact with these visualizations to understand their current status regarding algorithmic harms, which is indicated by triangle indicators, among the spectrum of values among all other users. Users can either select one of the three algorithmic harms by clicking the label, or activate the checkbox to see all algorithmic harms at once.

\section{Usage Scenario}
\label{sec:usage-scenario}

\begin{figure}
\includegraphics[width=\columnwidth]{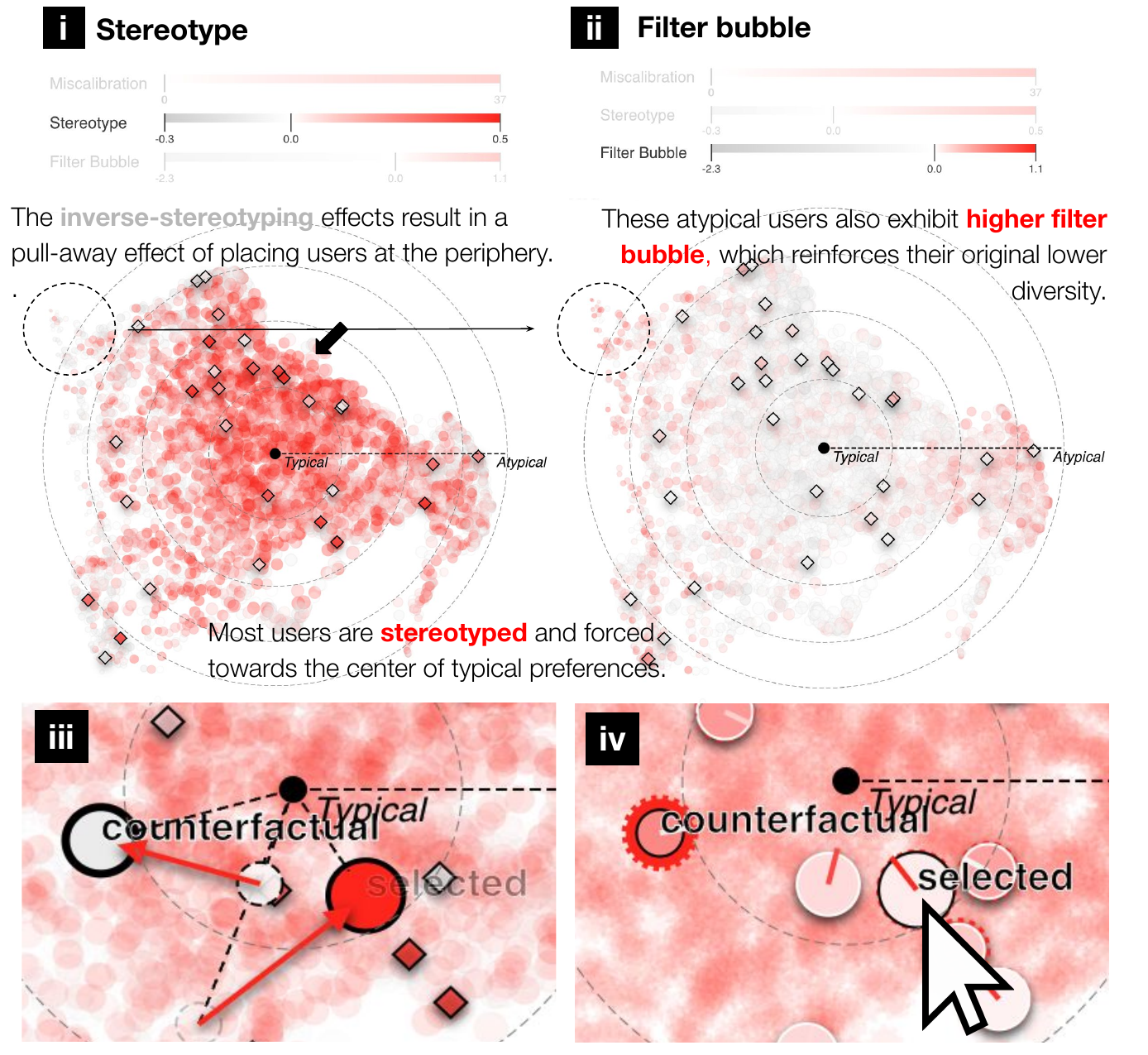}
    \vspace{-2em}
    \addtocounter{figure}{1}
    \caption{\label{fig:usage-scenario} (i-ii) Data practitioners' making sense of stereotype and filter bubble in reinforcing atypical users' narrower diversity. (iii-iv) End-users' comprehension process via counterfactual retrieval and prototypical cases.}
    \addtocounter{figure}{-2}
    \vspace{-2em}
\end{figure}

\subsection{Data practitioners' understanding of the penetration of algorithmic harms}
Consider Jane, a data practitioner working at a movie streaming company. She wanted to investigate how systematic biases affect users with different demographics and preferences, with a particular focus on understanding the disparities in the recommendations they receive.

Upon loading the interface, Jane first encountered the User Space, where users are positioned based on how much their preferences deviate from the mean. When she selected "stereotype" in the Algorithmic Harm Indicator, she immediately noticed that, while most of users exhibited a higher degree of stereotyping, atypical users positioned at the periphery with gray circles in this outer region (Fig. \ref{fig:usage-scenario}-i) were mostly gray, indicating that these users were subject to inverse-stereotyping, which pulled their preferences away from the mean.

Jane then examined the "filter bubble" effect for these same users (Fig. \ref{fig:usage-scenario}-ii). She observed that these users, who had lower diversity in their original preferences (as indicated by smaller circles), were also experiencing a higher degree of filter bubble in their recommendations as indicated by red-colored circles. This indicated that users who frequently engaged with specific genres were receiving increasingly homogeneous recommendations, which reinforced their existing views and contributed to the creation of a filter bubble.

The dashboard's visualizations provided Jane with a clear understanding of how these biases manifest, leading to significant insights. She realized that the system's current algorithm was not just marginalizing niche content but also reinforcing mainstream preferences, which could limit the diversity of content that users are exposed to. This understanding would allow her to make informed recommendations on how to adjust the algorithm to promote a more balanced and diverse set of recommendations, particularly for users whose preferences are not aligned with the majority.

\subsection{End-users' sense-making of algorithmic negative impact on their recommendations}
Consider Sarah, a frequent user of entertainment media who often felt that the recommendations she received did not align with her interests. She wished to understand the reasons behind this misalignment.

Upon loading the interface, Sarah first encountered her glyph in the User Space, where the algorithmic effects on her preferences were visualized. She immediately noticed that her glyph appeared closer to the typical preference. Selecting "stereotype" in the Algorithmic Harm Indicator and hovering over her icon (Fig. \ref{fig:teaser}C), she observed that her original preferences had deviated significantly from the typical ones. However, due to stereotyping in the algorithm, her preferences were forced toward the typical (Fig. \ref{fig:teaser}B-ii). 

By examining the prototypical user in her cluster, she discovered that this user exhibited higher levels of stereotyping and miscalibration, indicated by the dark red stereotyping bar on the glyph pointing towards the typical preference (Fig. \ref{fig:usage-scenario}-iv). In the user details (Fig. \ref{fig:teaser}A), she noticed that her preference for Romance was inflated (as indicated by the blue bar) and her preference for Horror was deflated (as indicated by the red bar). Curious about the impact of her demographics, Sarah used the Counterfactual Simulation to explore the query: "What if I were a man?" The system retrieved recommendations that a male user with similar overall preferences would receive. In the User Space, she found that, unlike her, the counterfactual user was visualized with an inverse-stereotyping effect (as indicated by a solid circle (Fig. \ref{fig:usage-scenario}-iii). Through these interactions, Sarah was able to understand that the recommender algorithm has a disparate impact from algorithmic effects that force some types of content to varying extents for different users.
\section{Conclusion}
\label{sec:conclusion}

In this study, we present a human-centric approach to addressing the transparency of algorithmic harms in recommender systems, with an interactive tool designed to help users make sense of them. A suite of visualizations, counterfactual explanations, and interactive modules, which were developed based on in-depth user perception interviews, enables users to explore the impacts of algorithmic harms on their recommendations. As a future work, we will focus on expanding the scope of our system to include more diverse datasets and enhanced data exploration capabilities as well as evaluate our tool with various types of users to understand the real-world impact of our tool.

\section{Acknowledgments}
The authors would like to acknowledge support from AFOSR, ONR, Minerva, NSF \#2318461, and Pitt Cyber Institute's PCAG awards. The research was partly supported by Pitt’s CRC resources (RRID:SCR 022735 through NIH \#S10OD028483). Any opinions, findings, and conclusions or recommendations expressed in this material do not necessarily reflect the views of the funding sources.


\bibliographystyle{abbrv-doi-hyperref}

\bibliography{references}


\end{document}